\documentclass[a4paper,11pt]{article}
\usepackage{amssymb}
\usepackage{amsfonts}
\usepackage{amsmath}
\usepackage{epsfig}

\begin{document}

\author{Duojie Jia$^{a,b\thanks{%
E-mail: jiadj@nwnu.edu.cn }}$, Yi-Ping Lu$^{a,b}$ \\
%EndAName
$^{a}$Institute of Theoretical Physics, College of Physics and \\
Electronic Engineering, Northwest Normal University,\textit{\ }\\
Lanzhou 730070, China,\\
$^{b}$Key Laboratory of Atomic and Molecular Physics \& Functional\\
Materials of Gansu Province, College of Physics and Electronic\\
Engineering, Northwest Normal University, Lanzhou 730070, China }
\title{A Ginzburg-Landau model of knotted flux-tube and glueball-like $f_{J}$
mesons}
\maketitle

\begin{abstract}
A dual Ginzburg-Landau model of the knotted chromo-electric flux-tube is
revisited, in which the covariant decomposition of gluon field and the
random phase approximation are used. It is shown that the SU(2) QCD vacuum
is of type-II superconductor, with the Ginzburg-Landau parameter $\kappa =%
\sqrt{3}$, being independent of the magnetic condensate and
strong coupling used, and consistent with the lattice data. The
mass spectrum of a number of $f_{J}$ meson states with $J\leq 2$,
which are taken to be of glue dominate, are computed with help of
the energies of the knotted(linked) QCD fluxtubes. The low-lying
$f_{J}$ states ($leq 1.7GeV$) are shown to be associated with the
string excitations of the types $2_{1}^{2}$ and $K=3_{1}^{{}}$ in
knot topologies.

PACS number(s): 12.38.-t, 11.15.Tk, 12.38.Aw

\end{abstract}

\section{Introduction}

According to the theory of strong interaction, quantum chromodynamics(QCD),
there will be pure-glue excitation known as glueball, and the $q\bar{q}$
states with explicit glue(known as hybrid mesons), which can have quantum
numbers forbidden to the $q\bar{q}$ systems in the naive quark model. These
gluonic excitations have been the subject of the extensive experimental and
theoretical studies(see, for instance,\cite{Klempt07,Crede09}) as the
identification and observation of these objects can be a good test of QCD at
the low-energy limit. After the early work of the bag-model\cite{JaffeBag}
of the glueballs, with mass prediction about $1.0$-$2.0GeV/c^{2}$, many
glueball (or, glue-rich meson) states were explored with various approaches,
including QCD on the lattice, the QCD sum rule, the constituent gluon
models, and etc., leading to a deep relation between the properties of these
states and the structure of the QCD vacuum\cite{Klempt07,Mathieu09}. Owing
to the mixing with the normal $q\bar{q}$ states, to accommodate these
gluonic excitations (except for the lightest tensor sector) remains to be an
open question\cite{Klempt07,Crede09,Mathieu09}.

The present work revisits the low-lying $f_{J}$ meson states (listed in the
Particle Data Group\cite{Hagiwara}) using the dual Ginzburg-Landau(GL) model
proposed in our previous work\cite{Jiadj08}, which is based on the
reformulated Yang-Mill(YM) theory in terms of the field decomposition\cite%
{Duan,Cho80,FN} of gluon variables. We show that the vacuum of the SU(2) QCD
is of a type-II superconductor type and the GL parameter is $\kappa =\sqrt{3}
$ for the vacuum condensate, being consistent with the lattice simulation%
\cite{DAlessandro07} and independent of the magnetic condensate as well as
the strong coupling that are used in the calculation. We further compute the
mass spectrum for a number of the low-lying $f_{J}$ meson states assuming
that the these mesons are mainly formed by the tightly knotted(or linked)
QCD fluxtubes, that is, the knotted(or linked) QCD strings (Hereafter, we
will also use the term knots for the closed types of strings including
links). The random vacuum approximation is used for the vacuum and main
features of a few low-lying $f_{J}$ states, the candidates for the
glue-dominated states, are examined in associated with the knot topology of
the chromo-electric fluxtubes for $J\leq 2$. Though the starting point for
model construction is the two-color gluodynamics, the relevance of the
present work to the real QCD can be inferred from the fact that $N$-color
QCD can be expressed as a sum over copies of two-color QCD\cite{Walker08} as
long as the Abelian dominance is valid for QCD\cite{tHooftB455}, as shown in
lattice QCD\cite{Bali,Ichie,Bissey}.

The idea of the gluonic fluxtubes\cite{Nielsen-Olesen,Nambu} can be traced
back to the early days of QCD when the resonance string(i.e., the QCD
string) was used to describe the confining force connecting the quarks\cite%
{Scherk}. The formation of a gluonic fluxtube between two widely separated
quarks is widely accepted, and is supported by the lattice QCD simulations%
\cite{Bali,Ichie,Bissey}. Meanwhile, it is expected that the glueballs, if
exist, are intimately related with the closed fluxtubes that can be omitted,
for instance, by a long linear string between the quarks. Using a flux-tube
model, Isgur and Paton\cite{Isgur} predicted that the lightest glueball has
the quantum number $J^{PC}=0^{++}$ and a mass of about $1.520GeV/c^{2}$. The
further predictions made later by the flux-tube model\cite{Iwasakietal} for
the three lightest glueball masses are consistent with the lattice
calculations\cite{Klempt07,Kronfeld}. Assuming the closed (knotted)
configurations for the gluonic fluxtubes, the glueballs were explored by
using the various actions, such as the Nambu-Goto action in the case of the
circular string\cite{Koma1999}, the nonlinear sigma actions \cite{Cho02} and
its extensions\cite{FNW2004} in the case of knotted strings. For the early
attempts to model hadrons in terms of closed strings of quantized fluxtube,
see Ref.\cite{Jehle}.

Our study is motivated by the ideal representation\cite{Katritch9697} of the
knot geometry, which provides a model-independent relationship between the
length-to-diameter ratio and the average crossing number of knot with the
given topology. This representation complements the nonlinear dynamics of
the stable knot found in field theory\cite{FN1997,Battye}, and is greatly
helpful in evaluating the glueball-like meson spectrum in terms of the
knotted objects in a bag model of fluxtube\cite{Buniy}. The calculations in
this work involve the Nielsen-Olesen(NO) vortices in the dual GL theory as
an effective description of the gluonic fluxtubes, for which the knot
geometry previously explored was utilized to calculate the energies of the
low-lying glueball-like meson states, similar to that in \cite{Buniy}.

\section{The order parameters in long-distance gluodynamics}

We begin with the reformulations of $SU(2)$ gluodynamics via the covariant
field decomposition of SU(2) gluon variables, known as
Cho-Faddeev-Niemi(CFN) decomposition\cite{Duan,Cho80,FN}. The gluon field $%
\vec{A}_{\mu }$ (the arrow denotes the three color indices $a=1,2,3$, along
the generators $\tau ^{a}$) is decomposed into \cite{Duan} $\vec{A}_{\mu
}=A_{\mu }\hat{n}+\vec{C}_{\mu }+\vec{X}_{\mu },$, in which $A_{\mu }=$ $%
\vec{A}_{\mu }\cdot \hat{n}$ is an Abelian gluonic potential, $\hat{n}(x)$
an unit isotriplet in color space, $\vec{C}_{\mu }:=$ $g^{-1}\partial _{\mu }%
\hat{n}\times \hat{n}$ the (non-Abelian) magnetic potential, and $\vec{X}%
_{\mu }$ (normal to $\hat{n}$) an covariant field, having to be constrained
by two extra conditions\cite{Shabanov}. We use the simplest choice of the
condition\cite{FN}, $\vec{X}_{\mu }=g^{-1}\phi _{1}\partial _{\mu }\hat{n}%
+g^{-1}\phi _{2}\partial _{\mu }\hat{n}\times \hat{n}$. With this change of
variables, the YM Lagrangian becomes
\begin{equation}
\mathfrak{L}^{YM}=-\frac{1}{4}[F_{\mu \nu }-\frac{Z(\phi )}{g}H_{\mu \nu
}]^{2}-\frac{1}{4g^{2}}\{(n_{\mu \nu }-iH_{\mu \nu })(\nabla ^{\mu }\phi
)^{\dag }\nabla ^{\nu }\phi +h.c\},  \label{dual}
\end{equation}%
where $F_{\mu \nu }\equiv \partial _{\mu }A_{v}-\partial _{v}A_{\mu }$, $%
Z(\phi )\equiv 1-|\phi |^{2}$, $\phi =\phi _{1}+i\phi _{2}$, $n_{\mu \nu
}\equiv \eta _{\mu \nu }(\partial \hat{n})^{2}-\partial _{\mu }\hat{n}\cdot
\partial _{\nu }\hat{n}$, $\nabla _{\mu }\phi \equiv (\partial _{\mu
}-igA_{\mu })\phi $ is the $U(1)$ covariant derivative induced by the gauge $%
U(1)$ rotation $U(\alpha \hat{n})=\exp (i\alpha n^{a}\tau ^{a})$ around
direction $\hat{n}$. The chromo-magnetic field $H_{\mu \nu }\equiv \hat{n}%
\cdot (\partial _{\mu }\hat{n}\times \partial _{\nu }\hat{n})$ is partially
dual to the chromo-electric field $F_{\mu \nu }$, as shown in \cite{Cho80,FN}%
. Such a change of variables in the reformulated dynamics (\ref{dual})
implements the Abelian projection \cite{tHooftB455} in a covariant way\cite%
{FN}. By choosing $\hat{n}(x)$ in (\ref{dual}) as an infrared order
parameter and using renormalization group analysis, the Skyrme-Faddeev(SF)
model\cite{FN1997}, which supports the knotted solitons with nonzero Hopf
charges, was proposed to model the low-energy dynamics of quantum YM\ theory%
\cite{FN}.

To see the implications of order parameter $\hat{n}(x)$, it is helpful to
re-examine the connection of the CFN field decomposition with the Abelian
projection from the viewpoint of the basis change. As shown in \cite%
{Shabanov} the choice of the constrains on $\vec{X}_{\mu }$ implies to
choose a local gauge transformation which maps the generic gluon field $\vec{%
A}_{\mu }$ into a gauge-fixed surface in the space of the gluon fields,
which yields that the map will necessarily be singular somewhere in
spacetime. The magnetic degree enters through the topological variable $\hat{%
n}(x)$, which provides the knot dynamics\cite{FN1997,Battye} and classifies
the physical region $V$ considered (as a mapping from $V\ $to $S^{2}$)
according to the homotopy $\pi _{2}(V)$. We note that when the $U(1)$
symmetry (rotation around $\hat{n}$ by angle $\alpha $) is left unbroken, $%
\hat{n}$ serves as the mapping transformation from the asymptotically free
gluon $\vec{A}_{\mu }$, which is represented in terms of the global basis \{$%
\tau ^{1\sim 3}$\}, to the infrared variables, which are represented in
terms of the local basis \{$\hat{n},\partial _{\mu }\hat{n}\mathbf{,}%
\partial _{\mu }\hat{n}\times \hat{n}$\}, with nontrivial metric in the
gauge group space. The QCD vacuum differs from the perturbative one owing to
the nontrivial homotopic class of the map $\hat{n}(x)$, or, equivalently, to
the singularities (magnetic charges) in the magnetic potential $\vec{C}_{\mu
}$, namely, the zeroes of the map $\hat{n}(x)$. The field decomposition in
terms of the local basis collapses when $\hat{n}$ becomes globally fixed $%
\hat{n}(x)\rightarrow \hat{n}_{0}$(i.e., the norm of $\partial _{\mu }\hat{n}
$ vanishes) so that \{$\hat{n},\partial _{\mu }\hat{n}\mathbf{,}\partial
_{\mu }\hat{n}\times \hat{n}$\} degenerates. In the latter situation, one
has instead to go back, discontinuously in the field mapping, to the usual
asymptotically free gluon variables $\mathbf{A}_{\mu }=A^{a}(x)\tau ^{a}$ in
the usual matrix representation of $\vec{A}_{\mu }$. This discontinuous
transition in the local gauge-fixing differs the asymptotically free phase
of QCD from the confining one. Thus, the nonvanishing vacuum expectation
value(VEV.) $\langle (\partial \hat{n})^{2}\rangle $ is required for the CFN
field decomposition to be a true change of variables.

Another order parameter arises by looking the confining phase of the theory (%
\ref{dual}). Taking the pure Abelian gauge $A_{\mu }=\partial _{\mu }\xi $
so that classically $F_{\mu \nu }=0$, as should be in the presumed vacuum
condensate, the theory (\ref{dual}) becomes
\begin{equation}
\begin{array}{c}
\mathfrak{L}^{M}=-\frac{Z^{2}(\rho )}{4g^{2}}H_{\mu \nu }^{2}-\frac{\rho }{%
g^{2}}[\partial ^{\mu }\rho \partial ^{\nu }s+g\partial ^{\mu }\xi \partial
^{\nu }\rho ]H_{\mu \nu } \\
+(2g^{2})^{-1}\left\{ [2g\partial ^{\mu }\xi \partial ^{\nu }s-g^{2}\partial
^{\mu }\xi \partial ^{\nu }\xi -\partial ^{\mu }s\partial ^{\nu }s]\rho
^{2}-\partial ^{\mu }\rho \partial ^{\nu }\rho \right\} n_{\mu \nu }%
\end{array}
\label{LM}
\end{equation}%
in which $\phi =\rho (x)e^{is(x)}$ has been used. One sees here that $Z(\rho
=|\phi |)$ resembles the dia-electric factor in the dia-electric soliton
model \cite{WIL89} and the gauge-invariant kernel in the effective model of
confinement \cite{tHooftA03} if one interprets the background media
associated with $\phi $ as the QCD vacuum. Explicitly, $Z(\phi \rightarrow
0)=1$ corresponds to the normal vacuum and $Z(\rho \rightarrow v)\neq 0$
(here, $v=\langle \rho \rangle $ is positive constant)$\,$to the condensate.
Indeed, taking the limit $\phi \rightarrow ve^{iF_{0}}$, the Eq. (\ref{LM})
can be further reduced to
\begin{equation}
\begin{array}{c}
\mathfrak{L}^{M}=-\frac{\langle Z^{2}(\rho )\rangle |_{\rho =v}}{4g^{2}}(%
\hat{n}\cdot \partial _{\mu }\hat{n}\mathbf{\times }\partial _{\nu }\hat{n}%
)^{2}-\frac{v^{2}(\partial \xi )^{2}}{2}(\partial _{\mu }\hat{n})^{2}+\frac{%
v^{2}}{2}\partial ^{\mu }\xi \partial ^{\nu }\xi (\partial _{\mu }\hat{n}%
\cdot \partial _{\nu }\hat{n}) \\
+V(n\cdot h,\xi )+\cdots .%
\end{array}
\label{GFN}
\end{equation}%
Here, the second term quadric in derivatives of $\hat{n}$ and potential
terms $V(n\cdot h,\xi )$ are added based on the renormalization group
analysis, as done in \cite{FN,FNW2004}. As we can see, (\ref{GFN}) is an
extended version of the SF model\cite{FN1997} (see also \cite{FNW2004}).
Given the nontrivial configuration $\hat{n}(x)$, the knot is classified by
non-zero Hopf charge $Q=1/(32\pi ^{2})\int d^{3}x\varepsilon
^{ijk}C_{i}H_{jk}$, with $C_{i}$ (having no local form in terms of $\hat{n}$%
) defined mathematically by $dC=H$ (The notation $H\equiv H_{jk}dx^{j}\wedge
dx^{k}/2=(\hat{n},d\hat{n}\wedge d\hat{n})$). Assuming there is a localized
excitation $G$ of $\hat{n}(x)$ in subregion $V_{G}\subset V$ and using the
virial theorem, the energy density estimated by (\ref{GFN}) is about $%
H_{G}\propto 2\langle (\partial _{\mu }\hat{n})^{2}\rangle +V_{\min }$,
which is in consistent with $\langle (\partial _{\mu }\hat{n})^{2}\rangle
\neq 0$ in $V_{G}$, while $\langle (\partial _{\mu }\hat{n})^{2}\rangle =0$
far away from $V_{G}$ due to the finite energy condition. The situation is
opposite for $\phi $, which tends to zero at the core of $V_{G}$ and to the
nonzero condensate $v=\langle \rho \rangle $ outside $V_{G}$, as it should
be according to the resemblance between $Z(\rho =|\phi |)$ in (\ref{LM}) and
the dia-electric factor.

The quanlitative behavior of $\hat{n}$ and $\phi $ in $G$ can be related to
the chromo-magnetic symmetry breaking, where the chromo-magnetic symmetry $%
H_{M}$ is defined by\cite{Cho80}%
\begin{equation}
D_{\mu }\hat{n}(x)\equiv (\partial _{\mu }+g\vec{A}_{\mu }\times )\hat{n}%
(x)=0.  \label{HM}
\end{equation}%
Observed that the CFN decomposition implies $D_{\mu }\hat{n}(x)=g(\vec{X}%
_{\mu }\times \hat{n})$, one sees that the breaking of $H_{M}$ is amount to
the nonvanishing VEV. of $(D_{\mu }\hat{n})^{2}$, namely,
\begin{eqnarray}
0 &\neq &g^{2}\langle (\vec{X}_{\mu }\times \hat{n})^{2}\rangle   \notag \\
&=&\langle \left( \phi _{1}\partial _{\mu }\hat{n}\times \hat{n}-\phi
_{2}\partial _{\mu }\hat{n}\right) ^{2}\rangle   \notag \\
&=&\langle \left( \phi _{1}^{2}+\phi _{2}^{2}\right) (\partial _{\mu }\hat{n}%
)^{2}\rangle ,  \label{HSB}
\end{eqnarray}%
in which the reparameterization \cite{FN} for $\vec{X}_{\mu }$ is used. In
the case of weak correlation between the off-diagonal variables ($\phi ,\hat{%
n}$), (\ref{HSB}) implies
\begin{equation}
\begin{array}{c}
0\neq \langle |\phi |^{2}\rangle \text{, Outside core}(V_{G})\text{,} \\
0\neq \langle (\partial _{\mu }\hat{n})^{2}\rangle \text{, in }V_{G}\text{,}%
\end{array}
\label{HOD}
\end{equation}%
which agrees with the existence of the order parameters $\hat{n}$ and $\phi $%
. In short, the arising of the order parameter $\hat{n}$ and $\phi $ are
associated with the breaking of $H_{M}$ defined by (\ref{HM}). Observed that
$(\vec{X}_{\mu }\times \hat{n})^{2}=\vec{X}_{\mu }^{2}$ in which $\vec{X}%
_{\mu }\cdot \hat{n}=0$, the chromo-magnetic symmetry breaking (\ref{HSB})
is amount to the $X$-field condensation%
\begin{equation*}
0\neq \langle (\vec{X}_{\mu })^{2}\rangle =\langle |\phi
|^{2}(g^{-1}\partial _{\mu }\hat{n})^{2}\rangle .
\end{equation*}%
Such type of condensation, referred as the off-diagonal gluon condensation,
was seen in the lattice simulations \cite{offconden} for gluodynamics.

\section{The random phase approximation}

To have the effective dynamics of the Abelian-Higgs multiplets ($A_{\mu
},\phi $), we need the further approximation associated with the vacuum of
QCD. We will use the random phase approximation(RPA), which can be due to
the residual $U(1)$ symmetry shared by the dual dynamics (\ref{dual}). To
see this, it is convenient to use two normalized basis $\{\hat{e}_{1},\hat{e}%
_{2}\}$, such that $\hat{e}_{1}\times \hat{e}_{2}=\hat{n}$, to describe the
2-dimensional plane $\Sigma _{(x)}$ which is normal to $\hat{n}(x)$ at $x$.
One can use a transformation from $\{\hat{e}_{1},\hat{e}_{2}\}$ to $\{\hat{e}%
_{+},\hat{e}_{-}\}$ explicitly for the isomorphic map: $SO(2)$ $\rightarrow
U(1)$.

Take $\hat{e}_{\mu }$ to be the direction vector of $\partial _{\mu }\hat{n}$%
, one can then write, as $\partial _{\mu }\hat{n}$ is in the plane $\Sigma
_{(x)}$,
\begin{equation}
\begin{array}{c}
\hat{e}_{\mu }=a_{\mu }^{1}\hat{e}_{1}+a_{\mu }^{2}\hat{e}_{2}, \\
=a_{\mu }\hat{e}_{+}+\bar{a}_{\mu }\hat{e}_{-},%
\end{array}
\label{ea}
\end{equation}%
where $a_{\mu }^{1,2}$are the two real four-vectors such that $(a_{\mu
}^{1})^{2}+(a_{\mu }^{2})^{2}=1$, and
\begin{equation*}
\begin{array}{c}
a_{\mu }=a_{\mu }^{1}-ia_{\mu }^{2},\bar{a}_{\mu }=a_{\mu }^{1}+ia_{\mu
}^{2}, \\
\hat{e}_{+}=\frac{1}{2}(\hat{e}_{1}+i\hat{e}_{2}),\hat{e}_{-}=\frac{1}{2}(%
\hat{e}_{1}-i\hat{e}_{2}).%
\end{array}%
\end{equation*}%
It is easy to show that
\begin{equation}
\begin{array}{c}
\hat{e}_{+}\cdot \hat{e}_{-}=1/2,\hat{e}_{+}\cdot \hat{e}_{+}=\hat{e}%
_{-}\cdot \hat{e}_{-}=0, \\
\hat{e}_{+}\times \hat{e}_{-}=-\frac{1}{2}i\hat{n}=-\hat{e}_{-}\times \hat{e}%
_{+}, \\
\hat{e}_{+}\times \hat{n}=i\hat{e}_{+},\hat{e}_{-}\times \hat{n}=-i\hat{e}%
_{-}, \\
\hat{e}_{\mu }{}^{2}=|a_{\mu }|^{2}=(a_{\mu }^{1})^{2}+(a_{\mu }^{2})^{2}=1.%
\end{array}
\label{formu}
\end{equation}%
The last equation of (\ref{formu}) implies that $a_{\mu }(x)=e^{i\theta
_{\mu }(x)}$, with $\theta _{\mu }(x)$ a real phase.

From the equations (\ref{formu}), one has
\begin{equation*}
\begin{array}{c}
\hat{e}_{\mu }\cdot \hat{e}_{\nu }=\frac{1}{2}(a_{\mu }\bar{a}_{\nu }+\bar{a}%
_{\mu }a_{\nu }) \\
=\cos (\theta _{\mu }-\theta _{\nu }),%
\end{array}%
\end{equation*}%
\begin{equation*}
\begin{array}{c}
(\hat{e}_{\mu }\times \hat{e}_{\nu })\cdot \hat{n}=\frac{i}{2}(\bar{a}_{\mu
}a_{\nu }-a_{\mu }\bar{a}_{\nu }) \\
=\sin (\theta _{\mu }-\theta _{\nu }),%
\end{array}%
\end{equation*}%
\begin{equation*}
\hat{e}_{\mu }\cdot \hat{e}_{\nu }+i(\hat{e}_{\mu }\times \hat{e}_{\nu
})\cdot \hat{n}=e^{i(\theta _{\mu }-\theta _{\nu })},
\end{equation*}%
Under the small gauge rotation $U(\alpha \hat{n})$, with $\alpha $ a small
angle, one can show that
\begin{equation*}
\begin{array}{c}
\hat{e}_{\mu }\rightarrow \hat{e}_{\mu }+\delta \hat{e}_{\mu }, \\
=\hat{e}_{\mu }+\hat{e}_{\mu }\times (\alpha \hat{n}), \\
=(a_{\mu }^{1}+\alpha a_{\mu }^{2})\hat{e}_{1}+(a_{\mu }^{2}-\alpha a_{\mu
}^{1})\hat{e}_{2}, \\
=a_{\mu }^{\prime }\hat{e}_{+}+\bar{a}_{\mu }^{\prime }\hat{e}_{-},%
\end{array}%
\end{equation*}%
where
\begin{equation*}
a_{\mu }^{\prime }=a_{\mu }+i\alpha a_{\mu }=e^{i\alpha }a_{\mu }.
\end{equation*}%
This means that the gauge rotation $U(\alpha \hat{n})$ corresponds to
\begin{equation}
\hat{e}_{\mu }\overset{U(\alpha \hat{n})}{\rightarrow }e^{i(\alpha +\theta
_{\mu })}\hat{e}_{+}+e^{-i(\alpha +\theta _{\mu })}\hat{e}_{-},
\label{shift}
\end{equation}%
or equivalently, to the phase shift $\theta _{\mu }\rightarrow \theta _{\mu
}+\alpha $.

By writing $\partial _{\mu }\hat{n}=M(x)\hat{e}_{\mu }$ and using (\ref{ea})
and (\ref{formu}), one has for the $SU(2)$ gluon field
\begin{equation}
\vec{A}_{\mu }=A_{\mu }\hat{n}+\frac{M(x)}{g}\left[ (i+\phi )a_{\mu }\hat{e}%
_{+}+h.c\right]  \label{A}
\end{equation}%
one can show that under $U(\alpha \hat{n})$
\begin{equation}
\begin{array}{c}
\vec{A}_{\mu }\rightarrow \vec{A}_{\mu }+\frac{1}{g}D_{\mu }(\vec{A}_{\mu
})(\alpha \hat{n}), \\
=\left( A_{\mu }+\frac{\partial _{\mu }\alpha }{g}\right) \hat{n}+\frac{M(x)%
}{g}\left[ (i+\phi +i\alpha \phi )a_{\mu }\hat{e}_{+}+h.c\right] , \\
=\left( A_{\mu }+\frac{\partial _{\mu }\alpha }{g}\right) \hat{n}+\frac{M(x)%
}{g}\left[ (i+e^{i\alpha }\phi )a_{\mu }\hat{e}_{+}+h.c\right] .%
\end{array}
\label{DA}
\end{equation}%
Comparison (\ref{A}) with (\ref{DA}) shows that the rotation $U(\alpha \hat{n%
})$ leads to transformation in the variables ($A_{\mu },\phi $) as exactly
as that for the Abelian Higgs multiplets: $A_{\mu }\rightarrow A_{\mu
}+\partial _{\mu }\alpha /g$, $\phi \rightarrow e^{i\alpha }\phi $. This
indicates that the unbroken $U(1)$ symmetry shared by the dual dynamics (\ref%
{dual}) is the local gauge rotation:$U(\alpha \hat{n})$.

We introduce the RPA such that $\langle \hat{e}_{\mu }\rangle =0$, which, by
noticing the residual symmetry (\ref{shift}), means
\begin{equation}
\langle e^{i\theta _{\mu }}\rangle =0\text{.}  \label{RPA}
\end{equation}%
Moreover, in the case that $\mu \neq \nu $ one has, from (\ref{RPA})
\begin{equation}
\langle e^{i(\theta _{\mu }-\theta _{\nu })}\rangle \simeq \langle
e^{i\theta _{\mu }}\rangle \langle e^{-i\theta _{\nu }}\rangle =0.
\label{theta}
\end{equation}%
The RPA introduced here assumes $\theta _{\mu }$ to be random distributed in
the QCD vacuum. As will be explored in the following section, the RPA is
very useful to find the effective $U(1)$ dynamics of the collective
variables ($A_{\mu },\phi $).

In order to describe the knot-like excitations, the SF-like dynamics of the
order parameter $\hat{n}$, similar to (\ref{GFN}), are usually used, as done
in \cite{Kondo,FNW2004}. Owing to the difficulty for extracting the
parameters in (\ref{GFN}), an alternative approach is to use the dynamics of
($A_{\mu },\phi $) to describe the fine profile of the chromo-electric
fluxtube, and to extract the energy of the closed fluxtube by utilizing the
values of the universal invariants for knot geometry \cite{Katritch9697},
which is main propose of this paper.

\section{The dual Ginzburg-Landau model}

From the section 2, we know that the order parameter $\phi (x)$ in (\ref%
{dual}) is suitable to play the role of soliton field interpolating in
between the two vacua: $\phi (x)=0$ and $\phi (x)=v(\neq 0)$. Writing $\phi
(x)=\Phi (x)+\delta \phi $, where $\Phi (x)$ is the complex condensate and $%
\delta \phi $ its quantum fluctuation, one has a nonzero correlation
\begin{equation}
\langle \phi (x)\phi ^{\dag }(y)\rangle \approx \Phi (x)\Phi ^{\ast }(y),%
\text{for }x^{0}>y^{0}.  \label{ass3}
\end{equation}

To find an effective model for ($A_{\mu },\phi $) starting from (\ref{dual}%
), let us take the mean field approximation
\begin{equation}
\partial _{\mu }\hat{n}(x)=M\hat{e}_{\mu }(x),  \label{PM}
\end{equation}%
with $\hat{e}_{\mu }$ the unit vectors defined in (\ref{ea}), and $M=\langle
(\partial _{\mu }\hat{n})^{2}\rangle ^{1/2}$ (no summing over $\mu $) a
constant in spacetime. The equation (\ref{PM}) yields
\begin{equation}
\begin{array}{c}
(\partial _{\mu }\hat{n})^{2}=M^{2}\{(\hat{e}_{0})^{2}-\sum_{i=1}^{3}(\hat{e}%
_{i})^{2}\}=-2M^{2} \\
H_{\mu \nu }=M^{2}h_{\mu \nu }\text{,} \\
h_{\mu \nu }=\hat{n}\cdot (\hat{e}_{\mu }\times \hat{e}_{\nu })=\sin \theta
_{\mu \nu }%
\end{array}
\label{nH}
\end{equation}%
with $\theta _{\mu \nu }=\theta _{\mu }-\theta _{\nu }$ the angle between
the directions $\hat{e}_{\mu }$\textbf{\ }and\textbf{\ }$\hat{e}_{\nu }$ in
the internal space. In addition, one has
\begin{equation}
\begin{array}{c}
\partial _{\mu }\hat{n}\cdot \partial _{\nu }\hat{n}=M^{2}\cos \theta _{\mu
\nu }, \\
n_{\mu \nu }=-M^{2}(2\eta _{\mu \nu }-\cos \theta _{\mu \nu }), \\
H_{\mu \nu }H^{\mu \nu }=M^{4}h_{\mu \nu }h^{\mu \nu }=\frac{M^{4}}{2}%
\sum_{\mu \nu }(1-\cos 2\theta _{\mu \nu }).%
\end{array}
\label{nn}
\end{equation}

Defining the magnetic condensate $H$ by $H^{2}\equiv \langle H_{\mu \nu
}H^{\mu \nu }\rangle $ and using the RPA, one finds, by using (\ref{nn}) and
(\ref{theta}),
\begin{equation}
H^{2}=6M^{4},  \label{H2}
\end{equation}%
namely, $M^{2}=H/\sqrt{6}$. In deriving (\ref{H2}), we have used $\sum_{\mu
\nu }1=12$, and $\langle \cos 2\theta _{\mu \nu }\rangle \simeq 0$ according
to the RPA. We see from (\ref{H2}) that the very existence of the
nonvanishing VEV. $\langle (\partial \hat{n})^{2}\rangle $ implies the
chromo-magnetic condensation, $\langle H_{\mu \nu }H^{\mu \nu }\rangle
\varpropto M^{4}$, which is due to the breaking (\ref{HSB}) of the magnetic
symmetry $H_{M}$.

It is important to note here that the RPA ignores the possible nontrivial
bending (or, twisting) of the $\hat{n}(x)$ orientation, which is crucial to
identify the knot-like excitations. In the case that $\hat{n}(x)$ $=\hat{n}%
_{cl}(x)+\delta \hat{n}$, where the variation of the classic part $\hat{n}%
_{cl}(x)$ is small(namely, the contribution to $M$ arises mainly from the
quantum fluctuation $\delta \hat{n}$), one can take $\hat{n}_{cl}$ to be
approximately parallel. Then, upon using \ref{RPA} and the mean field
approximation (\ref{PM}) as well as (\ref{nH}), the gluodynamics (\ref{dual}%
) becomes
\begin{eqnarray*}
\mathfrak{L}^{YM} &=&-\frac{1}{4}F_{\mu \nu }^{2}+\frac{M^{2}}{4g}Z(\phi
)F^{\mu \nu }h_{\mu \nu }-\frac{M^{4}}{4g^{2}}Z(\phi )^{2}h_{\mu \nu }^{2} \\
&&+\frac{M^{2}}{4g^{2}}\{[2\eta _{\mu \nu }+e^{i\theta _{\mu \nu }}](\nabla
^{\mu }\phi )^{\dag }\nabla ^{\nu }\phi +h.c\},
\end{eqnarray*}%
which, after utilizing (\ref{H2}), becomes
\begin{equation}
\mathfrak{L}^{eff}\simeq -\frac{1}{4}F_{\mu \nu }^{2}+\frac{H}{\sqrt{6}g^{2}}%
\langle (\nabla ^{\mu }\phi )^{\dag }\nabla ^{\nu }\phi \rangle -\frac{H^{2}%
}{4g^{2}}\langle Z(\phi )^{2}\rangle .  \label{LAH}
\end{equation}%
Here, the following relations, which are due to the RPA, are used, \
\begin{equation*}
\begin{array}{c}
\langle h_{\mu \nu }\rangle \simeq 0,\langle h_{\mu \nu }^{2}\rangle \simeq
6, \\
\langle e^{i\theta _{\mu \nu }}(\nabla ^{\mu }\phi )^{\dag }\nabla ^{\nu
}\phi \rangle \simeq 0.%
\end{array}%
\end{equation*}

Using the Wick theorem and the Bose symmetry of the scalar field, one can
show
\begin{eqnarray*}
\langle (\phi ^{\dag }\phi )^{2}\rangle &=&\langle \phi ^{\dag }\phi \rangle
\langle \phi ^{\dag }\phi \rangle +\langle \phi ^{\dag }\phi ^{\dag }\rangle
\langle \phi \phi \rangle +\langle \phi ^{\dag }\phi \rangle \langle \phi
^{\dag }\phi \rangle \\
&=&2\langle \phi ^{\dag }\phi \rangle ^{2},
\end{eqnarray*}%
and
\begin{equation*}
\begin{array}{c}
\langle Z(\phi )^{2}\rangle =\langle 1+(\phi ^{\dag }\phi )^{2}-2\phi ^{\dag
}\phi \rangle \\
\approx 1+2(\Phi ^{\ast }\Phi )^{2}-2\Phi ^{\ast }\Phi \\
=2(|\Phi |^{2}-1/2)^{2}+1/4.%
\end{array}%
\end{equation*}%
where (\ref{ass3}) is applied so that $\langle (\nabla ^{\mu }\phi )^{\dag
}\nabla _{\mu }\phi \rangle =\left( \nabla _{\mu }\Phi (x)\right) ^{\ast
}\nabla ^{\mu }\Phi (x)$. Rescaling the scalar $\Phi $ to that with
dimension of mass,
\begin{equation*}
\frac{\sqrt{H}}{\sqrt[4]{6}g}\Phi (x)\rightarrow \Phi (x),
\end{equation*}%
we obtain, from (\ref{LAH}), the effective dual GL model given by
\begin{equation}
\mathfrak{L}^{DGL}=-\frac{1}{4}F_{\mu \nu }^{2}+|(\partial _{\mu }-igA_{\mu
})\Phi |^{2}-V(\Phi )-\frac{H^{2}}{8g^{2}}.  \label{DGL}
\end{equation}%
In terms of the re-scaled complex $\Phi $, the potential in (\ref{DGL}) is
\begin{equation}
V(\Phi )=\frac{\lambda ^{2}}{4}(|\Phi |^{2}-v^{2})^{2},  \label{Vphi}
\end{equation}%
\ with the parameters given by
\begin{equation}
\begin{array}{c}
\lambda =\sqrt{12}g, \\
v=\frac{\sqrt{H}}{\sqrt[4]{24}g}.%
\end{array}
\label{ml}
\end{equation}

One sees that the effective model (\ref{DGL}) for the gluodynamics takes the
form of that for the dual superconductor\cite{Suzuki}. It is remarkable that
the scalar potential (\ref{Vphi}) assumes exactly the \textit{Mexico-hat}
form and ensures the dual Meissner effect for confining the chromo-electric
field $A_{\mu }$, since both of $H\sim M^{2}$ and $g$ are positive. We also
note that the ensuing dual superconductor picture with two vacua $\Phi =0$
and $\Phi =v$ in (\ref{DGL}) agrees with the vacuum picture discussed in
section 2.

It is known that the dual GL model (\ref{DGL}) admits the NO vortex solution
\cite{Nielsen-Olesen}, with two length scales: the coherent length $\xi
=1/m_{\Phi }$ and the penetrating length $\lambda _{L}=1/m_{A}$. The mass
scales $m_{\Phi }$ for the Higgs-like field $\Phi $ and $m_{A}$ for the
chromo-electric field $A_{\mu }$ are fixed by the explicit form of the
potential (\ref{Vphi}). Writing in terms of $\lambda $ and $v$ in (\ref{ml}%
), or, equivalently of the magnetic condensate $H$, they are
\begin{equation}
\begin{array}{c}
m_{\Phi }=\frac{\lambda }{\sqrt{2}}v=\frac{\sqrt[4]{6}}{\sqrt{2}}\sqrt{H},
\\
m_{A}=\sqrt{2}gv=\frac{\sqrt{H}}{\sqrt[4]{6}},%
\end{array}%
\text{ }  \label{mA}
\end{equation}%
which depend merely upon $H=\sqrt{6}\langle (\partial \hat{n})^{2}\rangle $
and are nonzero when the magnetic symmetry broken. Given the scales (\ref{mA}%
), one readily finds the GL parameter (defined by $\kappa =\lambda _{L}/\xi
) $ for the condensate vacuum to be \
\begin{equation}
\kappa =\frac{m_{\Phi }}{m_{A}}=\sqrt{3}\text{, (type-II),}  \label{para}
\end{equation}%
which is independent of the magnetic condensate $H$. The GL parameter given
by (\ref{para}) predicts the vacuum of the SU(2) gluodynamics to be of the
type of type-II superconductor, in nicely consistent with the lattice data $%
\kappa =\allowbreak 1.702$($=0.16fm/0.094fm$) for SU(2) gluodynamics\cite%
{DAlessandro07}. The very fact that the potential (\ref{Vphi}) in (\ref{DGL}%
) assumes the Mexico-hat form gives an independent argument for supporting
the dual superconductor mechanism of the low-energy phase of QCD proposed by
Nambu, 't\ Hooft and others\cite{Nambu,tHooftB455}. The type-II
superconductor was also confirmed in some of the lattice simulations, in
which\ $\kappa =1.\allowbreak 04$($=1.3614GeV/1.3123GeV$) \cite{Gubarev99}
and $\kappa =\allowbreak 1.49$($=0.164fm/0.11fm$)\cite{Alessandro2007} were
predicted.

\section{Glueball-like mesons as knotted fluxtubes}

Due to the $U(1)$ gauge symmetry for which $\pi _{2}(U(1)=1)$, the dual GL
model (\ref{DGL}) does not allow the stable solution of closed fluxtube
being against self-shrinking. To describe the knotted gluonic excitations,
it entails to have a nonlinear dynamics of knotted configuration, such as
SF-like model (\ref{GFN}), which is difficult to solve for now. A way out is
to use the Nambu-Goto action \cite{Koma1999} to prevent the self-shrinking
instability of the closed fluxtubes in the GL model. In this section, we use
(\ref{DGL}) as a model of the gluonic flux-tube profile and calculate the
glueball spectrum by taking into account both the string tension and the
twisting of the knotted fluxtubes, given that the geometric ratios of the
length to the diameter are known for a set of knot types\cite{Katritch9697}.

Consider a slice of the vortex cross-section, with the cylindrical
coordinates $(r,\theta )$ in it. We search for, as usual, the static NO(or,
the Abrikosov) vortex solution to the model (\ref{DGL}), in the Coulomb
gauge ($\mathbf{\nabla }\cdot \vec{A}=0$), in the form
\begin{equation*}
\begin{array}{c}
\vec{A}(\vec{x})=\hat{\theta}A(r)=-\frac{\hat{\theta}n}{gr}[1-F(r)],A_{0}=0,
\\
\Phi (\vec{x})=v\rho (r)\exp (in\theta ),%
\end{array}%
\end{equation*}%
with the boundary condition $F(r\rightarrow \infty )=0,\rho (r\rightarrow
\infty )=1$. The chromo-electric field is given by ($\vec{E},\vec{B}$), with
$\vec{E}=0$ in the present case, and
\begin{equation}
\vec{B}=\mathbf{\nabla }\times \vec{A}=\hat{z}\frac{n}{gr}\left( \frac{dF}{dr%
}\right) .  \label{BB}
\end{equation}

The static energy for (\ref{DGL}) is%
\begin{equation}
E^{DGL}=\int d^{3}x\left\{ \frac{\vec{B}^{2}}{2}+|(\partial
_{i}-igA_{i})\Phi |^{2}+V(\Phi )+\frac{H^{2}}{8g^{2}}\right\}  \label{DAH}
\end{equation}%
with $V(\Phi )$ given by (\ref{Vphi}). In terms of the vortex profiles ($%
F(r),\rho (r)$), the energy (\ref{DAH}) becomes
\begin{equation}
E=\int d^{3}x\left[ \frac{1}{2}\left( \frac{n}{gr}\right) ^{2}\left( \frac{dF%
}{dr}\right) ^{2}+v^{2}\left( \frac{d\rho }{dr}\right) ^{2}+\frac{n^{2}v^{2}%
}{r^{2}}F^{2}\rho ^{2}+V(\rho )\right] ,  \label{E}
\end{equation}%
where the constant energy ($\varepsilon \varpropto \int H^{2}d^{3}x$) in (%
\ref{DAH}) is treated as the zero-point energy. Using (\ref{Vphi}) and (\ref%
{ml}), one has for the potential,
\begin{equation*}
V(\rho )=\frac{H^{2}}{8g^{2}}\left( \rho ^{2}-1\right) ^{2}.
\end{equation*}

Introducing a dimensionless variable $x=r/a$, with $a$ a length scale, the
static equations of motion for (\ref{E}) are
\begin{equation}
\begin{split}
\frac{d^{2}F}{dx^{2}}-\frac{1}{x}\frac{dF}{dx}& =A\rho ^{2}F, \\
\frac{d^{2}\rho }{dx^{2}}+\frac{1}{x}\frac{d\rho }{dx}-\frac{n^{2}}{x^{2}}%
F^{2}\rho & =B\rho \left( \rho ^{2}-1\right) ,
\end{split}
\label{GLeq}
\end{equation}%
with the controlling parameters given by
\begin{equation}
\begin{array}{c}
A=(\sqrt{2}gva)^{2}, \\
B=(\lambda va/\sqrt{2})^{2}.%
\end{array}
\label{AB}
\end{equation}%
The asymptotic behavior of the vortex profiles $F(x)$ and $\rho (x)$ reads
\begin{equation}
\begin{array}{c}
F(x\rightarrow 0)\approx 1-c_{n}x^{2}, \\
\rho (x\rightarrow 0)=c_{n}^{\prime }K_{n}(\sqrt{B}x)\approx c_{n}^{\prime
}x^{n},%
\end{array}
\label{Asym1}
\end{equation}%
and
\begin{equation}
\begin{array}{c}
F(x\rightarrow \infty )=C_{n}xK_{1}(\sqrt{A}x)\approx C_{n}\sqrt{x}e^{-\sqrt{%
A}x}, \\
\rho (x\rightarrow \infty )=1-C_{n}^{\prime }K_{0}(\sqrt{2B}x)\approx 1-%
\frac{C_{n}^{\prime }}{\sqrt{x}}e^{-\sqrt{2B}x},%
\end{array}
\label{Asym2}
\end{equation}%
with $c_{n}$,$c_{n}^{\prime }$,$C_{n}$,$C_{n}^{\prime }$ the constants for
given $n$, and $K_{j}(x)$ the Bessel function of second type. The
expressions (\ref{Asym1}) and (\ref{Asym2}) are useful and will be
incorporated in the specification of the boundary condition in the process
of the numerical relaxation for solving (\ref{GLeq}) in finite (but large)
interval.

Assuming that the vortex has a finite length $L$ along $z$ direction and
integrating the angular($\theta $) part, one has, from (\ref{E})
\begin{equation}
E^{L}=2\pi L\left[ \frac{(n/g)^{2}}{a^{2}}I^{glue}+v^{2}I^{B}+\frac{%
a^{2}H^{2}}{g^{2}}I^{N}\right] ,  \label{knotE}
\end{equation}%
with three dimensionless integrals given by
\begin{equation}
\begin{array}{c}
I^{glue}=\int_{0}^{\infty }\frac{dx}{2x}\left( \frac{dF}{dx}\right) ^{2}, \\
I^{B}=\int_{0}^{\infty }dx\left[ x\left( \frac{d\rho }{dx}\right) ^{2}+\frac{%
n^{2}}{x}F^{2}\rho ^{2}\right] , \\
I^{N}=\int_{0}^{\infty }dx\left[ \frac{x}{8}\left( \rho ^{2}-1\right) ^{2}%
\right] .%
\end{array}
\label{II}
\end{equation}%
The first term in (\ref{knotE}) is the chromo-electric energy and roughly
scales as $\sim L(tr\Phi _{E}^{2})/(\pi a^{2})$, due to the flux
conservation, with $\Phi _{E}$ the flux of chromo-electric field (\ref{BB})
in the area $\pi a^{2}$ of the vortex cross-section. Since the three
integrals in (\ref{II}) are all positive for the NO vortex, the stable
vortex exists, for which the first and third terms in (\ref{knotE}) balance.

The minimization of (\ref{knotE}) with respect to $a$ yields%
\begin{equation}
a(=a_{n})=\frac{k_{n}}{\sqrt{H}},  \label{a}
\end{equation}%
which depends explicitly on the vortex quantum number $n$. In deriving (\ref%
{a}), (\ref{ml}) are used. On the other hand, using (\ref{ml}), (\ref{AB})
and (\ref{a}), one has
\begin{equation}
A=\frac{k_{n}^{2}}{\sqrt{6}},B=\frac{\sqrt{6}k_{n}^{2}}{2}.  \label{ABk}
\end{equation}%
This means that the energy depends on $a_{n}$ not only through (\ref{knotE})
explicitly, but also implicitly through the integrals (\ref{II}), which rely
on $a_{n}$ through the controlling parameters $A$ and $B$(both of them $\sim
a_{n}^{2}$).

For given input ($\lambda ,v$), or equivalently, ($g,H$), we numerically fix
the the length scale $a$ in the case $n=1\sim 7$, in two steps: (1) to find
the vortex profile ($F,\rho $) by solving the GL equation (\ref{GLeq}) using
relaxation method with $A$ and $B$ given by (\ref{AB}) for a given $a$; (2)
to find the scale $a=a_{n}$ for which the string tension $\sigma =E^{L}/L$
is minimized by calculating $E^{L}/L$ as a function of $a$ through (\ref%
{knotE}) and (\ref{II}) for the profile given in (1). The optimal $a$ and
 the corresponding $k_{n}$, including three integrals ($I^{glue},I^{B},I^{N}$%
), are listed in $\mathrm{Table}$\textrm{\ I}. The numerical results are
shown in \textrm{FIG.1 } for the profile ($F,\rho $) and \textrm{FIG.2 }for
the tension $\sigma (a)$ as a function of $a$, for $n$ up to $4$. The
relation (\ref{ABk}) are found to be fulfilled for calculated $a_{n}$ in (%
\ref{a}). In the process of the numerical relaxation, the asymptotic
behavior (\ref{Asym1}) and (\ref{Asym2}) are incorporated in the boundary
condition and the initial vortex profiles $F_{initial}(x)=\sec h(\sqrt{A}x)$
and $\rho _{initial}(x)=\tanh (\sqrt{2B}x)$ are used.

%%%%%%%%%%%%%%%%%%%%%%%%%%%%%%%%%%%%%%%%%%%%%%%%%%%%%%%%%%%%%%%%%%%%%%%%%%%%%%%%%%%%%%%%%
\begin{tabular}[t]{|c|c|c|c|c|c|c|c|}
\hline $n$ & $1$ & $2$ & $3$ & $4$ & $5$ & $6$ & $7$ \\
\hline
$k_{n}$ & 1.0019 &  1.2982 &  1.5204 & 1.5945 & 1.7426 & 1.9648 &  2.2611 \\
\hline
$r_{n}(Gev^{-1})$ & 0.5012 & 1.0490 & 1.4235 & 1.6156 & 1.8551 & 2.1420 & 2.5520 \\
\hline
$I^{glue}$ & 0.1941 & 0.1941 & 0.1942 & 0.1942 & 0.1936 & 0.1933 & 0.1944 \\
\hline
$I^{B}$ & 0.6770 & 1.5893 & 2.8129 & 4.3547 & 5.7913 & 6.5896 & 7.3093 \\
\hline
$I^{N}$ & 0.0370 & 0.0418 & 0.0461 & 0.0490 & 0.0525 & 0.0601 & 0.0687 \\
\hline
$a_{n}(Gev^{-1})$ & 1.2850 & 1.6650 & 1.9500 & 2.0450 & 2.2350 & 2.5200 & 2.9000 \\
\hline
\multicolumn{8}{c}{Table I} \\
\end{tabular}

%%%%%%%%%%%%%%%%%%%%%%%%%%%%%%%%%%%%%%%%%%%%%%%%%%%%%%%%%%%%%%%%%%%%%%%%%%%%%%%%%%%%%%%%
\begin{figure}[h]
\begin{center}
\rotatebox{0}{\resizebox *{8.5cm}{6.7cm} {\includegraphics
{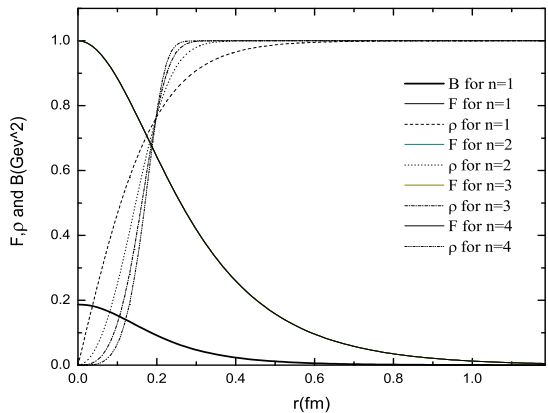}}}
\end{center}
\caption{The vortex profiles for $n=1\sim 4$. $F$ decreases with
$r$ and changes slightly for different $n$, while $\protect\rho $
increases faster
as $n$ increases. The plot is also given for the chromo-electric field $B$%
(with unit of $GeV^{2}$), which decreases from $0.1866Gev^{2}$(at $r=0$) to $%
0$(at $r=\infty $). Here, the strong coupling and the magnetic
condensation are taken to be $g=3.2198$, $H=0.5246GeV^{2}$. }
\end{figure}

%%%%%%%%%%%%%%%%%%%%%%%%%%%%%%%%%%%%%%%%%%%%%%%%%%%%%%%%%%%%%%%%%%%%%%%%%%%%%%%%%%%%%%%
\begin{figure}[h]
\begin{center}
\rotatebox{0}{\resizebox *{8.5cm}{6.7cm}
{\includegraphics{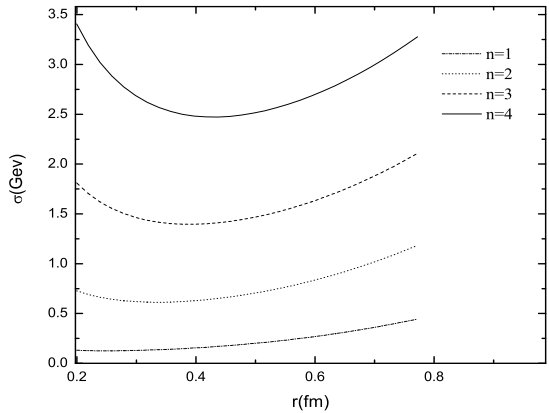}}}
\end{center}
\caption{The string(vortex) tension $\protect\sigma =E^{T}/L$ as
a function
of the length scale $a$ for $n=1\sim 4$. Here, $g=3.2198$, $H=0.5246GeV^{2}$%
. }
\end{figure}

It can be seen that (\ref{knotE}) describes the linear rising ($\sim \sigma
L $) of the confining potential for large $L$. When $L$ decreases, there
will be corrections to (\ref{knotE}) due to short-range gluonic
interactions. In QCD we can expect, quite generally, the full energy for the
low-lying spectrum of the confining fluxtube with the fixed ends at distance
$L$ to be\cite{Luscher,Kuti99}%
\begin{equation}
E(L)=\sigma L+\varepsilon +\frac{c_{0}}{L}+\frac{m\pi }{L}+\mathcal{O}(\frac{%
1}{L^{2}}).  \label{EL}
\end{equation}%
Here, $\sigma $ is the string tension, $\varepsilon $ the the zero-point
energy in (\ref{DAH}), $m\pi /L$ ($m=1,2,\cdots $) the vibrationally-excited
energies of the string flux, and $c_{0}=-\pi /12$ is the Casimir energy of
zero-point fluctuations of the string\cite{Luscher}. Combining (\ref{EL})
with (\ref{knotE}), one obtains for the energy of the knot-like gluonic
fluxtube.
\begin{equation}
E_{n,m}=4\pi e(K)r_{n}\left[ \frac{(n/g)^{2}}{a_{n}}I^{glue}+a_{n}v^{2}I^{B}+%
\frac{a_{n}^{3}H^{2}}{g^{2}}I^{N}\right] +\varepsilon +\frac{c_{0}+m\pi }{%
2e(K)r_{n}a_{n}},  \label{em}
\end{equation}%
where $r_{n}=R/a_{n}$ is the ratio of the core radius $R$ of the vortex to $%
a_{n}$ in (\ref{a}), and $e(K)=L/(2R)$ a topological invariant which is
universal for a given knot types\cite{Katritch9697}. The factor $r_{n}$
depends merely upon $n$ and can be fixed through the numerical relaxation
for (\ref{GLeq}). In principle, $e(K)$ can be evaluated by the dynamics of
knot like (\ref{GFN}). This is not available by now since we fail to fix the
explicit form of the potential $V(n\cdot h,\xi )$ in (\ref{GFN}).
Fortunately, $e(K)$ were previously determined for a set of knot types ($%
K^{\prime }s$) via the Monte-Carlo simulations\cite{Katritch9697},
partial of which are listed in $\mathrm{Table}$\textrm{\ II} and
\textrm{\ III} for our references.

We use the fluxtube energy (\ref{em}) for these knot types(denoted by $   %
K=2_{1}^{2},3_{1}$, etc. in topology) to model the spectrum of the
glueball-like mesons. The computed spectra are shown and compared
with the data of the partial $f_{J}$ states ($J\leq 2$) (also a
few $\eta$ states) in the $\mathrm{Table}$ \textrm{\ II } in the
case of $n=1$ and of some low $m$'s. In the evaluation of the
glueball-like meson energies using (\ref{em}), the strong coupling
and chromo-magnetic condensate are chosen as

\begin{equation}
g=3.2198,H=0.5246GeV^{2}  \label{P1}
\end{equation}%
so that $4\alpha _{s}/3=g^{2}/(3\pi )=1.1$ (see Ref.\cite{Isgur}) and
\begin{equation}
\begin{array}{c}
|\epsilon _{Vac}|=H^{2}/(8g^{2})=(0.24Gev)^{4}; \\
v=\sqrt{H/2}/(\sqrt[4]{6}g)=0.1016GeV%
\end{array}
\label{Param}
\end{equation}%
with $\epsilon _{Vac}$ the vacuum energy density, given in the lattice
computation\cite{Chernodub2004}. $H$ is fixed based on (\ref{mA}) through
the lattice data for the coherent length $\xi =1/m_{\Phi }$ and penetrating
length $\lambda _{L}=1/m_{A}$. The numeric result for $a_{n}$ is $%
a_{1}=1.2850$ for $n=1$ with the tension
\begin{equation*}
\sigma =0.125GeV^{2},
\end{equation*}%
and the other cases $n\geq 2$ are excluded since they yield anomalously high
string tension, such as $\sigma =0.339$, $0.695$, etc., see \textrm{FIG.2}.

Different with (\ref{P1}), we alternatively choose, by taking $v=0.1094GeV$
and $\epsilon _{Vac}=-0.7GeV$, the input values of the parameters which
change $H$ slightly, and thereby the scalar condensate $v=\langle \Phi
\rangle $ (see (\ref{ml})) so that
\begin{equation}
g=3.2198,H=0.6080GeV.  \label{P2}
\end{equation}%
The numerical data corresponding to (\ref{P2}) are shown in Table
\textrm{II} and \textrm{III } for the mass spectrum of a number of
low-lying $f_{J}$ states with $J\leq 2$.

It can be seen from $\mathrm{Table}$\textrm{\ II} and \textrm{Table III }%
that a remarkable agreement was reached between the experimental and
predicted spectrum for the meson $f_{J}$ states, showing that the most of
the meson $f_{J}$ states can be identified as the knotted string excitations
of the types $(n,m)=(1,j),j=0,1,2$. A little different suggestion for these
identification are given in $\mathrm{Table}$\textrm{\ IV}, with the
prediction for glueballs with quantum number $J^{CP}=0^{++},0^{-+},0^{+-}$,
compared to the lattice data.

We account for the results in $\mathrm{Table}$\textrm{\ II} and \textrm{III }%
as follows: a number of the meson $f_{J}$ states, when taken to
be the glueball-like (i.e., the glueball dominate) states, can be %
viewed as a knot excitation of the chromo-electric fluxtubes with
knot types $n_{1}^{l}$($ n=1\sim 5,l=1,2$) in the topology terminology,%
in which the low-lying ($\leq 1.71GeV/c^{2}$)glueball-like meson states %
are best described by the fluxtube of the knot types $2_{1}^{2}$ and $3_{1}^{{}}$,%
with $ (n,m)=(1,j),j=0,1,2$. It is unknown in our framework why some of%
vibrational modes lack the experimental counterparts, but it is known that%
the knot types $K=2_{1}^{2},3_{1}^{{}}$ corresponds to the low-lying $f_{J}$ %
states because they have the simplest topologies among the knot types $%
2_{k}^{l},3_{k}^{l}$, in the sense that they are non-shrinkable
and they are shortest in the unit of the string diameter. While
the minor variation, roughly $200MeV$ per unit of $m$, in the
spectrum arises from the vibrational modes of the fluxtube, the
spectrum of glueball-like mesons are mainly due to the knot
geometry of fluxtubes.

\begin{tabular}[t]{lllllll}
\hline\hline States & Mass$^{a}$ & $m$ & $K^{b}$ & $e(K)^{c}$ &
$E_{n,m}(G)$ &  \\ \hline $f_{0}(600)$ & 400-1200 & $1$ & $2_{1}$
& 6.2832 & 674.6 &  \\ \hline $f_{0}(980)$ & 980$\pm$10 & $0$ &
$2_{1}^{2}$ & 12.6 & 976.0 &  \\ \hline $h_{1}(1170)$ &  & $2$ &
$2_{1}$ & 6.2832 & 1246.9 &  \\ \hline $a_{1}(1260)$ &
$$
& $ 1$ & $2_{1}^{2}$ & 12.6 & 1262.2 &  \\ \hline $f_{0}(1370)$ &
1200-1500 & $1$ & $2_{1}^{2}$ & 12.6 & 1262.2 &  \\ \hline
$f_{0}(1500)$ & 1507$\pm$5 & $0$ & $3_{1}$ & 16.4 & 1500.1 &  \\
\hline $f_{0}(1710)$ & 1718$\pm$6 & $1$ & $3_{1}$ & 16.4 & 1719.4
&
\\ \hline
$\eta(1760)$ & 1760$\pm$11 & $3$ & $2_{1}^{2}$ & 12.5664 & 1757.0 &  \\
\hline $\eta(2225)$ & 2220$\pm$18 & $1$ & $2_{1}^{2}*0_{1}$ &
20.8496 & 2278.4 &
\\ \hline
$f_{2}(1910)$ & 1915$\pm$7 & $2$ & $3_{1}$ & 16.4 & 1938.7 &  \\
\hline
$f_{2}(2010)$ & $2011^{+60}$ & $4$ & $2_{1}^{2}$ & 12.5664 & 2015.5 &  \\
\hline $f_{0}(2100)$ & 2103$\pm$7 & $3$ & $3_{1}$ & 16.4000 &
2100.4 &  \\ \hline $f_{2}(2150)$ & 2156$\pm$11 & $0$ & $4_{1}$ &
21.2000 & 2153.6 &  \\ \hline
$f_{0}(2200)$ & 2189$\pm$13 & $0$ & $4_{1}^{2} $ & 21.4000 & 2180.8 &  \\
\hline $f_{2}(2300)$ & 2297$\pm$28 & $4 $ & $3_{1}$ & 16.4000 &
2298.5 &  \\ \hline
$f_{2}(2340)$ & 2339$\pm$60 & $1 $ & $4_{1}^{2}$ & 21.4000 & 2348.8 &  \\
\hline
\multicolumn{6}{c}{Table II} &  \\
\end{tabular}

\begin{tabular}[t]{lllllll}
\hline\hline States & Mass$^{a}$ & $m$ & $K^{b}$ & $e(K)^{c}$ &
$E_{n,m}(G)$ &  \\ \hline $f_{0}(600)$ & 400-1200 & $1$ & $2_{1}$
& 6.2832 & 674.6 &  \\ \hline $f_{0}(980)$ & 980$\pm$10 & $0$ &
$2_{1}^{2}$ & 12.6 & 976.0 &  \\ \hline $h_{1}(1170)$ &  & $2$ &
$2_{1}$ & 6.2832 & 1246.9 &  \\ \hline $a_{1}(1260)$ &
$$
& $ 1$ & $2_{1}^{2}$ & 12.6 & 1262.2 &  \\ \hline $f_{0}(1370)$ &
1200-1500 & $1$ & $2_{1}^{2}$ & 12.6 & 1262.2 &  \\ \hline
$f_{0}(1500)$ & 1507$\pm$5 & $0$ & $3_{1}$ & 16.4 & 1500.1 &  \\
\hline $f_{0}(1710)$ & 1718$\pm$6 & $1$ & $3_{1}$ & 16.4 & 1719.4
&
\\ \hline
$\eta(1760)$ & 1760$\pm$11 & $3$ & $2_{1}^{2}$ & 12.5664 & 1757.0 &  \\
\hline $\eta(2225)$ & 2220$\pm$18 & $1$ & $2_{1}^{2}*0_{1}$ &
20.8496 & 2278.4 &
\\ \hline
$f_{2}(1910)$ & 1915$\pm$7 & $2$ & $3_{1}$ & 16.4 & 1938.7 &  \\
\hline
$f_{2}(2010)$ & $2011^{+60}$ & $4$ & $2_{1}^{2}$ & 12.5664 & 2015.5 &  \\
\hline $f_{0}(2100)$ & 2103$\pm$7 & $3$ & $3_{1}$ & 16.4000 &
2100.4 &  \\ \hline $f_{2}(2150)$ & 2156$\pm$11 & $0$ & $4_{1}$ &
21.2000 & 2153.6 &  \\ \hline
$f_{0}(2200)$ & 2189$\pm$13 & $0$ & $4_{1}^{2} $ & 21.4000 & 2180.8 &  \\
\hline $f_{2}(2300)$ & 2297$\pm$28 & $4 $ & $3_{1}$ & 16.4000 &
2298.5 &  \\ \hline
$f_{2}(2340)$ & 2339$\pm$60 & $1 $ & $4_{1}^{2}$ & 21.4000 & 2348.8 &  \\
\hline
\multicolumn{6}{c}{Table III} &  \\
\end{tabular}

$^{a}$ The data from the PDG summary tables and\cite{Crede09}.
$^{b}$ The knot types in terms of notation $n_{k}^{l}$, means a
link of $l$ components
with $n$ crossing, and occurring in the standard table of links() on the $k$%
th place. $^{c}$ The data from the simulations\cite{Katritch9697}
except for $2_{1}^{2}$ and $2_{1}^{2}\ast 0_{1}$. $^{d}$ The data
from Lattice simulations.

\begin{tabular}[t]{lllllll}
\hline\hline States & Mass$^{d}$ & $m$ & $K^{b}$ & $e(K)^{c}$ &
$E_{n,m}(G)$ &  \\ \hline $0^{++}$ & 1710$\pm$130 & $1$ & $3_{1}$
& 16.4 & 1678.5 &  \\ \hline $0^{-+}$ & 2560$\pm$155 & $0$ &
$5_{1}$ & 24.2000 & 2561.1 &  \\ \hline $0^{+-}$ & 4780$\pm$290 &
$0$ & $9_{2} $ & 40 & 4703.3 &  \\ \hline
\multicolumn{6}{c}{Table IV} &  \\
&  &  &  &  &  &
\end{tabular}

\section{Summary and discussions }

The dual dynamics of the SU(2) QCD is revisited using the covariant
(Cho-Faddeev-Niemi) decomposition of the gluon field. Assuming that the
chromo-magnetic symmetry is broken and using the random phase approximation,
we show by deriving a dual Ginzburg-Landau model for SU(2) gluodynamics that
the QCD vacuum is of type-II superconductor. The Ginzburg-Landau parameter
is shown to be $\kappa =\sqrt{3}$, which is independent of the
chromo-magnetic condensate $H$ and the strong coupling $g_{s}$ used, and
agrees well with the lattice simulation. The mass spectrum of a number of
low-lying $f_{J}$ states are calculated for the chromo-flux quantum $n=1$
and found to be in a good agreement with the recent lattice data, by adding
the energy correction arising from the twisting of the knotted(linked) QCD
fluxtubes and the zero-point energy to the energy of the Nielsen-Olesen
vortex in the dual Ginzburg-Landau model and applying the universal
topological invariants of the knot geometry being the length-to-diameter
ratios of the fluxtubes. A number of meson $f_{J}$ states are identified as
the knot-like gluonic excitations in the form of closed chromo-electric
fluxtubes with the chromo-flux quantum $n=1$ and the vibrationally-excited
string mode $m=2$. The most of the low-lying $f_{J}$ states are found to be
described by the knotted string excitations of the types $2_{1}^{2}$ and $%
K=3_{1}^{{}}$ topologies, having the quantum number $(n,m)=(1,j),j=0,1,2$.

Given that the glueballs can mix with the quark-antiquark states,
as proposed in \cite{Amsler}, our calculation present a support
that some of the $q\bar{q}$-states(mainly $f_{J}$ states and a
few $\eta $), listed in \textrm{Table II }and \textrm{III},can
have a dominate component of glues, namely, the energy of the
valence quarks is negligible, and as a result, these $f_{J}$
states can be well described by the knotted fluxtubes formed by
the collective gluons. The origin of the uncertainty in the
knot-type identifications for a few states, such as $f_{0}(980)$ and $%
f_{0}(1370)$, $f_{2}(2220)$, $\eta (2225)$ in \textrm{Table II
and III}, remains unclear yet, and the further studies are needed
for the origin though the quark hybrid component of the states is
probable origin.

D. Jia is grateful to X. Liu for discussions. This work is supported in part
by National Natural Science Foundation of China (No.10965005), The
Project-sponsored by SRF for ROCS, SEM, and by The Project of Key Laboratory
of Atomic and Molecular Physics \& Functional Materials of Gansu Province.

\end{document}